\documentclass[aps,reprint,showpacs,preprintnumbers,amsmath,amssymb,prb]{revtex4-1} 
\usepackage{graphicx}
\graphicspath{{figs/}{./}}

\usepackage{color}
\newcommand\FigureFile[1] {#1.eps}
\DeclareGraphicsExtensions{pdf}

\usepackage{dcolumn}% Align table columns on decimal point
\usepackage{bm}% bold math
\usepackage{color}

\newcommand\eq[1]                              % single unlabeled equation
{
\begin{align*}
#1
\end{align*}
}

\newcommand\eql[2] % single labeled equation
{
\begin{equation}\label{#1}
\begin{split}
#2
\end{split}
\end{equation}
}

\newcommand\eqsl[1]                            % multiply labeled equation
{
\begin{align}
#1
\end{align}
}

\newcommand\eqssl[2]                      % multiply labeled SUB-equations
{
\begin{subequations}\label{#1}
\begin{align}
#2
\end{align}
\end{subequations}
}

\newcommand\Eq[1]      {Eq.~\eqref{#1}}
\newcommand\Eqs[1]     {Eqs.~\eqref{#1}}
\newcommand\Fig[1]     {Fig.~\ref{#1}}
\newcommand\Sec[1]     {Sec.~\ref{#1}}

\newcommand\Ref[1]     {Ref.~\onlinecite{#1}}
\newcommand\Refs[1]    {Refs.~\onlinecite{#1}}
\newcommand\ket[1]     {|{{#1}}\rangle}

\newcommand\PsiGS      {\Psi_0}
\newcommand\PsiT[1][]  {\Psi_{\mathrm{T}#1}^{}}
\newcommand\Dc[1]      {c_{{#1}}^{}}
\newcommand\Cc[1]      {c_{{#1}}^\dagger}
\newcommand\Half       {\frac{1}{2}}

\newcommand\rvec       {\mathbf{r}}

\newcommand\Hop        {{\hat{H}}}
\newcommand\Kop        {{\hat{K}}}

\newcommand\Vop        {{\hat{V}}}
\newcommand\vop        {{\hat{v}}}
\newcommand\Ebind      {E_{\mathrm{b}}}
\newcommand\EbindHF    {E_{\mathrm{b}}^{\mathrm{HF}}}
\newcommand\Ebindcorr  {E_{\mathrm{b}}^{\mathrm{corr}}}
\newcommand\Efrag      {E_{\mathrm{frag}}}
\newcommand\EHF        {E_{\mathrm{HF}}}
\newcommand\Ecorr      {E_{\mathrm{corr}}}

\newcommand\Eh[1][]    {\ensuremath{E_\mathrm{h}#1}}
\newcommand\mEh[1][]   {\textrm{m}\ensuremath{E_\mathrm{h}#1}}

\newcommand\Order[1]   {\mathcal{O}\left(#1\right)}

\newcommand\Htwo       {\textrm{H$_{\textrm{2}}$}}
\newcommand\Caplus     {\textrm{Ca\textsuperscript{+}}}
\newcommand\CaH        {\textrm{Ca\textsuperscript{+}\textendash\,\Htwo}}
\newcommand\CaHHHH     {\textrm{Ca\textsuperscript{+}\textendash\,4\Htwo}}
\newcommand\dHH        {\ensuremath{d_{\textrm{H}-\textrm{H}}}}

\definecolor{xmgrace-green4}{rgb}{0.0,0.55,0.0}
\definecolor{Green}{rgb}{0.2,0.96,0.2}
\definecolor{Remarks}{rgb}{1,0.3,0.3}%red
\definecolor{Extra}{rgb}{0.2,0.2,1}%blue
\definecolor{Blue}{rgb}{0.2,0.3,1}
\definecolor{Black}{rgb}{0,0,0}

\newcommand\COMMENTED[1] {}

\begin{document}

% Try a shorter title:
\title{
Assessing weak hydrogen binding on
Ca\textsuperscript{+} centers:
%for hydrogen storage:
An accurate many-body study with large basis sets
}

\author{Wirawan Purwanto}
\email{wirawan0@gmail.com}
\affiliation{Department of Physics, College of William and Mary,
Williamsburg, Virginia 23187-8795, USA}

\author{Henry Krakauer}
\affiliation{Department of Physics, College of William and Mary,
Williamsburg, Virginia 23187-8795, USA}

\author{Yudistira Virgus}
\affiliation{Department of Physics, College of William and Mary,
Williamsburg, Virginia 23187-8795, USA}

\author{Shiwei Zhang}
\affiliation{Department of Physics, College of William and Mary,
Williamsburg, Virginia 23187-8795, USA}

\date{\today}

\begin{abstract}

Weak {\Htwo} physisorption energies present a significant challenge
to even the best correlated theoretical many-body methods.
We use the phaseless auxiliary-field
quantum Monte Carlo (AFQMC) method
to accurately predict the binding
energy of  {\CaHHHH}. Attention has recently focused on this model chemistry
to test the reliability of electronic structure methods for {\Htwo}  binding on dispersed alkaline
earth metal centers.
A modified Cholesky decomposition
is implemented to realize
the Hubbard-Stratonovich transformation efficiently
with large Gaussian basis sets.
We employ the largest correlation-consistent Gaussian type basis sets available,
up to cc-pCV5Z for Ca,
to accurately extrapolate to the complete basis limit.
The calculated potential energy curve exhibits
binding with a double-well structure.

\end{abstract}

% PACS, the Physics and Astronomy Classification Scheme.
%Use showkeys class option if keyword display desired

\maketitle

\section{Introduction}
\label{sec:intro}

%\REMARKS{motivate the Ca+4H2 system}

Clean energy economy
has become an appealing worldwide endeavor because of its
promise of environmental friendliness and economic security.
Developing improved hydrogen fuel storage systems for fuel cell applications and finding new technologies for production of hydrogen from renewable sources
are important components in achieving this goal.
%Hydrogen molecule (\Htwo) has long been recognized
%as a potential alternative for clean energy application.
A major standing challenge
%to the realization of hydrogen-based energy economy
is the lack of a method for efficient hydrogen storage and retrieval.
In developing technologies for on-board hydrogen storage systems,
the U.S.~DOE has set a target capacity of $9$ wt $\%$ by
year 2015.\cite{DOE-H2-FCTP}
Physisorption of hydrogen molecules is one mechanism under
consideration for use in hydrogen storage.
For operation at ambient temperatures,
suitable storage
media should have a binding strength
of \mbox{$\sim 0.1 - 0.4$ eV} per {\Htwo}.\cite{DOE-H2-2010-report,Bhatia2006}
However, our current understanding of physisorption processes is still
lacking.\cite{DOE-H2-2010-report}
% see http://www.hydrogen.energy.gov/pdfs/progress10/iv_c_1a_simpson.pdf
First-principles calculations could potentially play an important role
in providing insight into the physics of physisorption
and identifying suitable media and mechanisms.
Achieving predictive accuracy for such weakly bound
systems has been problematic, however, for the
usually successful density functional theory (DFT)
approach and even for
explicitly correlated methods.

Dispersed alkaline-earth metal systems have recently
shown some promise as {\Htwo}  storage media.\cite{Sun2009,Kim2009,Lee2009,Cha2009,Ohk2010,Sun2010,Bajdich2010,carbyne-2011}
Attention has focused on a simple model chemistry
to test the reliability of electronic structure methods in predicting the
binding of  {\Htwo} on  {\Caplus} centers.
(This is a simplified model which does not take into account
the zero-point motion and entropy effects, which are important in modeling
real hydrogen storage.)
A common thread from these calculations is the prediction and characterization of a double-well
potential energy curve (PEC) in the symmetric dissociation of {\CaHHHH}.
An outer van der Waals well is found in some of the correlated calculations, but not in the DFT calculations with
standard local or semilocal exchange-correlation functionals.\cite{Cha2009,Ohk2010,Sun2010,Bajdich2010}
An inner well
(Kubas complex \cite{Kubas2001,Kubas2007}) is found at shorter distances.
Overbinding is found in local and semilocal DFT calculations,
while explicitly correlated methods yield conflicting results regarding the magnitude of the binding or whether
this inner well is even bound. \cite{Cha2009,Ohk2010,Bajdich2010}
Disagreement between the many-body calculations can be partly attributed to inadequate basis set convergence.
The earliest calculation, using the second-order M{\o}ller-Plesset perturbation
method (MP2), predicted an inner local minimum
that is not bound. \cite{Cha2009}
Ohk {\it et al.} \cite{Ohk2010} subsequently
used MP2 with larger basis sets
with an added correction from
coupled cluster with singles and doubles and perturbative triples
[CCSD(T)]
to find both bound inner and outer wells.
As we will show in \Sec{sec:results}, however,
the basis sets in \Ref{Ohk2010} were
still too small for accurate extrapolation to the complete basis set (CBS) limit,
due to neglect of (semi)core-valence correlation effects.
The most recent many-body calculation
used diffusion Monte Carlo (DMC)
to find an unbound inner well and
a barely bound outer well. \cite{Bajdich2010}
The current status of theoretical work is clearly unsatisfactory,
and the reasons for this need to be clarified and remedied.

The primary objectives of our work are to produce an accurate
first-principle PEC of the {\CaHHHH} symmetric dissociation process
and to clarify the key factors that control the accuracy for many-body calculations.
The phaseless auxiliary-field quantum Monte Carlo (AFQMC) method \cite{Zhang2003, Purwanto2004, Purwanto2005, AlSaidi2006b,Purwanto2009_Si}
with standardized Gaussian type orbital (GTO) basis,\cite{EMSL_BasisSets2007}
hereafter
designated as GTO-AFQMC,\cite{Zhang2003,AlSaidi2006b}
is used to compute the PEC and to
carefully extrapolate the result to the
CBS limit.
AFQMC has been applied to study
a wide variety of material systems.
\cite{Zhang2003,Purwanto2004,Purwanto2005,AlSaidi2006a,AlSaidi2006b,
AlSaidi2006c,AlSaidi2007,AlSaidi2007b,Suewattana2007,Purwanto2008,
Purwanto2009_C2,Purwanto2009_Si}
Its accuracy
has been shown to be similar to the gold-standard
CCSD(T) method
near equilibrium geometries and better
for bond-breaking.
\cite{AlSaidi2006b,AlSaidi2007b,Purwanto2008,Purwanto2009_C2}
The key features of AFQMC include:
(1) low algebraic scaling with the system size [$\Order{M^3-M^4}$,
where $M$ is the number of the one-particle basis functions];
(2) ease of implementation on massively parallel supercomputers;
%and
(3) its wide applicability as an {\it ab-initio} method in condensed matter physics
and quantum chemistry.
In the context of quantum chemistry, GTO-AFQMC uses the
\emph{identical} basis and Hamiltonian
of other standard methods,
such as Hartree-Fock (HF), MP2, CCSD(T), and
configuration interaction (CI).

In this paper we implement a significant technical improvement
in GTO-AFQMC, which removes a computational bottleneck in the preprocessing and initialization step.
This allow us to use very large basis sets
to obtain accurate ground state energies at the CBS limit.
This is made possible by the implementation of
a modified Cholesky decomposition (mCD)\cite{Beebe1977,Koch2003,MOLCAS7}
of the two-body interaction term in the Hamiltonian.
Our GTO-AFQMC calculation with the aug-cc-pCV5Z basis set (827 GTOs)
represents the largest
many-body calculation in this system with GTOs
to date.

The rest of this paper is organized as follows.
The implementation of mCD together with
relevant methodological details of AFQMC are
presented in \Sec{sec:method}.
Accuracy and timing illustrations of AFQMC/mCD are also given.
GTO-AFQMC {\CaHHHH} PEC calculations and extrapolation to the CBS limit
are presented in
\Sec{sec:results}.
In \Sec{sec:discuss} we
discuss possible residual sources of error and compare our results to previously published results.
We summarize and conclude in \Sec{sec:summary}.

\section{GTO-AFQMC methodological improvements and calculation details}
\label{sec:method}

After a brief AFQMC overview, methodological improvements with GTOs are presented. The improvements remove
a preprocessing bottleneck for computer time and memory, allowing the use of large GTO basis sets
in AFQMC. This is achieved in an unbiased manner using
mCD.
Accuracy and timing illustrations are given, and the section concludes
with the calculation details which are used in \Sec{sec:results}.

\subsection{AFQMC ground state projection}

We briefly review the key features of the phaseless AFQMC method, which are relevant for the remainder of this section.
Detailed descriptions of the method can be found in
\Refs{Zhang2003, Purwanto2004, Purwanto2005, AlSaidi2006b,
Purwanto2009_Si}.

AFQMC stochastically evaluates the ground state of a many-body
Hamiltonian $\Hop$
by means of importance sampled random walks in Slater-determinant space.
\cite{Zhang1997_CPMC,Zhang2003}
The stochastically generated determinants are the samples of the
ground-state many-body wave function $\ket{\PsiGS}$.
Although AFQMC is in principle an exact method,
in practice the sign or complex-phase problem arises,
\cite{Loh1990,Zhang1997_CPMC,Zhang2003}
which must be controlled using some approximate means
in order to prevent the exponential growth of the Monte Carlo variance.
This is done with the phaseless approximation,
\cite{Zhang2003}
which has been demonstrated to yield excellent agreement with
both experimental and known exact results
in a wide variety of molecules and extended systems.
\cite{Zhang2003, % original phaseless
AlSaidi2006a,    % TiO/MnO
AlSaidi2006b,    % GAFQMC
AlSaidi2006c,    % post-d
AlSaidi2007,     % H-bonded
AlSaidi2007b,    % bondbreaking
Suewattana2007,  % PWQMC
Purwanto2008,    % F2
Purwanto2009_C2, % C2
Purwanto2009_Si}
To date we have developed AFQMC methods with two
basis sets:
(1) planewaves with pseudopotentials, which are suitable for studying periodic systems;
\cite{Zhang2003, % original phaseless
Zhang2005,       % CPC paper
AlSaidi2006a,    % TiO/MnO
AlSaidi2007,     % H-bonded
Suewattana2007,  % PWQMC
Purwanto2009_Si} % Si
%and
(2) GTOs, which are widely employed
for \emph{ab initio} quantum chemistry. %for isolated molecules.
\cite{AlSaidi2006b,    % GAFQMC
AlSaidi2006c,    % post-d
AlSaidi2007,     % H-bonded
AlSaidi2007b,    % bondbreaking
Purwanto2008,    % F2
Purwanto2009_C2} % C2
In addition there have been many applications on
lattice models ({\it e.g.}, the Hubbard model)
where the phaseless approximation reduces to the constrained path approximation
\cite{Zhang1997_CPMC,ChangPRB2008} due to the nature of the interaction.
These systems typically have correlation energy as a significantly higher fraction of their
total energy than in most molecules and solids. The high accuracy that has been
achieved in these systems for both energy and correlation functions is thus encouraging
for the prospect of AFQMC in real materials.

The electronic Hamiltonian
$
    \Hop = \Kop + \Vop
$
contains one-body
%\eql{eq:Kop}
%{
$
    \Kop
  = \sum_{ij} K_{ij} \Cc{i} \Dc{j}
$
%}
kinetic energy and external potential terms plus
two-body electron-electron interaction terms,
\eql{eq:Vop}
{
    \Vop
& = \Half \sum_{ijkl} V_{ijkl} \Cc{i} \Cc{j} \Dc{k} \Dc{l}
    \,,
}
expressed here in a second quantized form.
The fermionic creation operators $ \Cc{i} $
are defined on a finite set of $M$ orthonormal
one-particle basis functions $\{ \chi_i(\rvec) \}$.
%The indices $i$, $j$, $k$, $l$ refer to these basis functions.

The AFQMC method projects the ground state wave function $\ket{\PsiGS}$
from a trial wave function $\ket{\PsiT}$,
\eql{eq:gs-proj}
{
    e^{-\tau \Hop}
    e^{-\tau \Hop}
    \cdots
    e^{-\tau \Hop}
    \ket{\PsiT}
  \rightarrow \ket{\PsiGS}
    \,,
}
using a short-time Trotter-Suzuki decomposition
\eql{eq:Trotter}
{
    e^{-\tau \Hop}
  = e^{-\tau \Kop / 2}
    e^{-\tau \Vop}
    e^{-\tau \Kop / 2}
  + \Order{\tau^3}
    \,.
}
The input $\PsiT$ can be either a single Slater determinant or
a multi-determinant wave function;
the AFQMC projection yields a stochastic multi-determinant representation
of $\PsiGS$.

The one-body projector is straightforward
to implement in AFQMC:
$e^{-\tau \Kop / 2}$ acting on a Slater determinant yields
another Slater determinant.
%Evaluation of the two-body projector $e^{-\tau \Vop}$
%ee employ the Hubbard-Stratonovich (HS) transformation
%\cite{Stratonovich1957,Hubbard1959}
%as the auxiliary-field transformation in \Eq{eq:HS-generic}.
To evaluate the two-body projector $e^{-\tau \Vop}$,
we first decompose $\Vop$ into a sum of squares of one-body operators,
\cite{Zhang2000_BookChapter}
\eql{eq:Vop-decomp}
{
    \Vop
%  = -\Half \sum_{\gamma=1}^{2 M^2}
  = -\Half \sum_\gamma
    \vop_\gamma^2
  + \textrm{(one-body term)}
    \,,
}
where the minus sign is just a notational convention, and
additional one-body terms
can arise from reordering the creation and destruction operators.
We employ the Hubbard-Stratonovich (HS) transformation
\cite{Stratonovich1957,Hubbard1959}
to rewrite the two-body projector as a multi-dimensional integral,
\eql{eq:HS-xform-noFB}
{
    %e^{-\tau \Vop}
    e^{(1/2) \tau \sum_\gamma \vop_\gamma^2}
    \underset{\tau \to 0}{=}
    \prod_\gamma \int_{-\infty}^{\infty} \frac{d\sigma_\gamma}{\sqrt{2\pi}}
    e^{-\sigma_\gamma^2 / 2}
    e^{\sqrt{\tau} \, \sigma_\gamma \vop_\gamma}
    \, .
}
This integral over $\{\sigma_\gamma\}$ is evaluated stochastically
in AFQMC.
Both \Eqs{eq:Trotter} and \eqref{eq:HS-xform-noFB} are exact
in the limit \mbox{$\tau \to 0$}.
We therefore have an exact reformulation of the
original ground-state projector
in terms of one-body projection operators $\{\vop_\gamma\}$
coupled with external auxiliary fields $\{\sigma_\gamma\}$
which, after integration,
recovers the original two-body interactions.
\cite{Zhang2000_BookChapter}

\subsection{Modified Cholesky decomposition}

\subsubsection{A potential computational bottleneck}

The HS decomposition  [\Eq{eq:Vop-decomp}] of the two-body interaction term
is a preprocessing step, which is done only once at the beginning of the calculation.
The decomposition is not unique, and
this flexibility can be exploited to obtain better performance and/or
accuracy of the AFQMC calculation.
\cite{Zhang2000_BookChapter}
The two-body interaction matrix elements $V_{ijkl}$ in Eq.~(\ref{eq:Vop}) are
given by
electron-electron Coulomb repulsion integrals (ERIs) in electronic structure
calculations:
\eql{eq:Vmtx}
{
   V_{ijkl}
 = \int d\rvec_1 d\rvec_2 \,
    \chi_i^*(\rvec_1) \chi_j^*(\rvec_2)
    %V(\rvec_1, \rvec_2)
    \frac{e^2}{|\rvec_1 - \rvec_2|}
    \chi_l(\rvec_1) \chi_k(\rvec_2)
    \,.
}
In our earlier implementation of GTO-AFQMC,\cite{AlSaidi2006b}
we used the straightforward approach of diagonalizing the
$M^2 \times M^2$ symmetric ERI supermatrix
$
    V_{\mu(i,l), \nu(j,k)}
$,
where
$\mu \equiv (i,l)$ and $\nu \equiv (j,k)$ are compound indices.
(The dimension $M^2$ is reduced by a factor of two with GTOs as they
are real-valued.)
With the
eigenvalues $\lambda_\gamma$
and eigenvectors $X^\gamma_{\mu(i,l)}$
of the  real, symmetric $V_{\mu\nu}$ supermatrix,
the HS one-body
operators can be written as
\eql{eq:vop-diag}
{
    \vop_\gamma
  = \sqrt{-\lambda_\gamma}
    \sum_{i,l}
    X^\gamma_{\mu(i,l)} \Cc{i} \Dc{l}
    \,.
}
In practice, only $\mathcal{O}(M)$ of the eigenvalues are found to have magnitudes greater than  $\sim 10^{-8} \, \Eh$,
so the remainder %(and their corresponding eigenvectors)
can be discarded.
The direct diagonalization leads to an
$\mathcal{O}(M^6)$ scaling of computer time
and $\mathcal{O}(M^4)$ storage.
While exact for any
Hermitian two-body interaction,
this approach clearly scales poorly with system size.

The HS decomposition described above is
a bottleneck for treating large systems.
For special choices of the basis, such as planewaves, or for model systems, such as the on-site Hubbard model, the
two-body interaction is easily written into the bilinear
HS form of \Eq{eq:Vop-decomp}, with $\mathcal{O}(M)$ terms. This and the diagonalization results above suggest that
the information content in the two-body term is only $\mathcal{O}(M)$ and that more efficient HS strategies could be devised for general basis
sets in electronic structure calculations.

The underlying problem here is the sheer size of the two-body supermatrix,
which plagues all {\it ab initio\/} quantum chemistry methods.
A number of approaches have been
devised to reduce the computer time and
number of integrals that need to be stored.
These include density fitting %(DF)
or other auxiliary basis methods,
\cite{Whitten1973,Vahtras1993,Weigend2002,Werner2003,Neese2003}
resolution of Coulomb operator, %(RC),
\cite{Varganov2008,Limpanuparb2009}
and 
mCD.
\cite{Beebe1977,Koch2003,MOLCAS7}
%These techniques reduce the overall computational scaling by
%providing various approximate ways to compute the ERIs
%so as to achieve the desired calculation accuracy.

We have chosen to implement the mCD to carry out the HS decomposition in AFQMC.
The method, similar to planewave approaches, has a
single threshold parameter  $\delta$ which determines the maximum error
in the Cholesky-expanded ERIs. For symmetric, positive semidefinite $V_{\mu\nu}$ supermatrices,
this error can be reduced to zero, within machine precision, for sufficiently small $\delta$.
In practice, our calculations use $\delta$ in the range $10^{-6}$ to $10^{-4} \Eh$, depending the target statistical accuracy in the AFQMC calculation, resulting
in $N_\mathrm{CD} \lesssim 7.5 M$ Cholesky vectors. The algorithm and its performance are
discussed next.

\subsubsection{Implementation of Cholesky decomposition in AFQMC}
\label{sec:cholesky}

The symmetric, positive semidefinite ERI supermatrix $V_{\mu\nu}$ is decomposed
using a recursive mCD algorithm.
\cite{Koch2003,MOLCAS7}
Given $J$ Cholesky vectors $L_{\mu}^{j}$ ($j=1 ... J$), $V_{\mu\nu}$ can be expressed as
\eql{eq:Vmtx-chol-def}
{
    V_{\mu\nu}
%& = \sum_{\gamma = 1}^{J}
%    L_{\mu}^{\gamma}
%    L_{\nu}^{\gamma}
& = \sum_{j = 1}^{J}
    L_{\mu}^{j}
    L_{\nu}^{j}
  + \Delta_{\mu\nu}^{(J)}
\\
&   \equiv
    V_{\mu\nu}^{(J)}
  + \Delta_{\mu\nu}^{(J)}
    \,,
}
where $\Delta_{\mu\nu}^{(J)}$ is the residual error at the $J$-th iteration.
The $(J+1)$-th Cholesky vector is obtained from
\eql{eq:L-new}
{
    L_{\mu}^{J+1}
  = \frac{\Delta_{\mu[\nu]_J}^{(J)}}
         {\sqrt{\Delta_{[\nu]_J[\nu]_J}^{(J)}}}
    \,,
}
where $[\nu]_J$ indicates the index of the largest diagonal element,
$\Delta_{[\nu]_J[\nu]_J}^{(J)}$, of the residual error matrix in the
$J$-th iteration.
A key point is that only a single column $\Delta_{\mu[\nu]_J}^{(J)}$ need
to be computed and stored at any iteration.
This iteration is repeated until
$\Delta_{[\nu]_J[\nu]_J}^{(J)}$ is less than $\delta$. This
procedure guarantees that all
matrix elements of the residual matrix are less than
$\delta$:
\eql{eq:V-chol-cond}
{
    \left|
        V_{\mu\nu}
      - V_{\mu\nu}^{(N_\mathrm{CD})}
    \right|
 =  \left|
        \Delta_{\mu\nu}^{(N_\mathrm{CD})}
    \right|
    \le
    \delta
    \,
}
where the number of Cholesky vectors corresponding to this $\delta$
is denoted by
$N_\mathrm{CD}$.
Damped prescreening \cite{MOLCAS7} is applied
in each iteration step
to stabilize the mCD factorization.
%(This prescreening automatically includes the initial prescreening
%procedure mentioned in \Ref{Koch2003}.)

Using the $N_\mathrm{CD}$ Cholesky vectors, the decomposition in \Eq{eq:Vop-decomp}
becomes
\eql{eq:Vop-chol-decomp}
{
    \Vop
& = \Half
    \sum_{\gamma=1}^{N_\mathrm{CD}}
    \left(
        \sum_{il} L_{\mu(i,l)}^\gamma \Cc{i}\Dc{l}
    \right)
    \left(
        \sum_{jk} L_{\nu(j,k)}^\gamma \Cc{j}\Dc{k}
    \right)
\\
& - \,
    \Half
    \sum_{\gamma=1}^{N_\mathrm{CD}}
    \sum_{ijk}
        L_{\mu(i,j)}^\gamma L_{\nu(j,k)}^\gamma \Cc{i}\Dc{k}
\\
& + \,
    \Order{\delta}
    \,,
}
where the extra one-body
operator is also explicitly expressed in terms of the Cholesky vectors.
The Cholesky vectors translate directly to the matrix elements of
$\vop_\gamma$ in \Eq{eq:Vop-decomp}:
\eql{eq:vop-chol}
{
%    \vop_\gamma^{\textrm{cholesky}}
    \vop_\gamma
  = \sqrt{-1}
    \sum_{il} L_{\mu(i,l)}^\gamma \Cc{i}\Dc{l}
    \,,
}
and the number of auxiliary fields
is just given by
$N_\mathrm{CD}$.

In this work
the Cholesky vectors were generated ``on the fly" within
NWChem\cite{NWChem-6.0}
or
MPQC.\cite{MPQC-2.3.1}
In GTO-AFQMC,
the second-quantized expression for the Hamiltonian must be expressed
with respect to orthogonalized GTOs. \cite{AlSaidi2006b}
The mCD procedure [\Eq{eq:Vmtx-chol-def}]
is carried out in the original GTO basis, and
the resulting Cholesky vectors are then transformed to the
orthogonalized basis.

% COMMENT ON THE ACCURACY OF ORIGINAL ERI
% THIS IS YUDIS' FINDING
% AT ISSUE: matrix elements must be accurately computed
% or else the Cholesky vectors are garbage
%
In order to produce high-quality Cholesky vectors which satisfy the
accuracy condition \Eq{eq:V-chol-cond},
%for all $(\mu, \nu)$ pairs,
it is imperative that the original ERIs used in the mCD %iteration
recursive procedure
be calculated to very high accuracy and precision.
%Numerical errors in the calculated ERIs leads to
%a
The $V_{\mu\nu}$ supermatrix will not be
strictly positive semidefinite (within machine precision)
if there are errors in the calculated ERIs.
%The effect of the small numerical errors in $V_{\mu\nu}$ would be
%magnified as the size the system ($M$) grows.
In a test case with $M = 180$, we observe from
direct diagonalization of $V_{\mu\nu}$
that there are some negative eigenvalues
of $\Order{-10^{-8}}$.
In this case, the resulting Cholesky vectors would not properly
reconstitute all the ERIs to within $\delta$ accuracy
when $\delta$ is driven below $10^{-8}$.
In
another test
with $M \sim 550$,
errors in the calculated ERIs
smaller than $10^{-8} \Eh$
lead to
violation of \Eq{eq:V-chol-cond} with
$\delta$ set to $\Order{10^{-6}}$.

\subsubsection{GTO-AFQMC/mCD accuracy and timing illustrations}

\newcommand\EDD {\ensuremath{E_{\mathrm{DD}}}}
\newcommand\EmCD {\ensuremath{E_{\mathrm{mCD}}}}

\begin{figure}[htbp]
\includegraphics[scale=0.31]{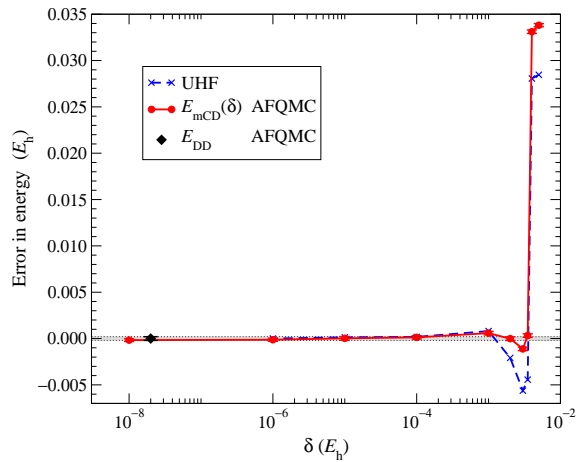}
\caption{\label{fig:AFQMC-chol-cutoff}
(Color online)
AFQMC total energy error, $\EmCD(\delta) - \EDD$,
as a function of the mCD threshold parameter $\delta$, where
$\EDD$ is the energy obtained from direct diagonalization of the
Coulomb matrix (energies are given in Table~\ref{tbl:AFQMC-chol-cutoff}).
The width of the gray line indicates the statistical uncertainty of $\EDD$.
%Coulomb matrix (energies are given in Table~I). % DIRTY HACK
The corresponding error (with respect to the UHF energy) of the trial wave function variational energy is also shown.
}
\end{figure}

Table~\ref{tbl:AFQMC-chol-cutoff}
compares the accuracy of GTO-AFQMC calculations using mCD with
that using direct diagonalization (DD) of the $V_{\mu\nu}$ supermatrix.
For each $\delta$ shown, we confirmed that all
$V_{\mu\nu}$ matrix elements are
reproduced with error less than $\delta$.
Within statistical uncertainty, the QMC energies are equivalent
whether we use DD or mCD
with $\delta$ ranging from $10^{-8}$ % {\Eh}
through $10^{-3}$ {\Eh}.
For this system, the truncation bias from mCD exceeds the targeted statistical error of $2\times 10^{-4}$ {\Eh}
when $\delta \gtrsim 3\times 10^{-3}$ {\Eh}.
\begin{table}[htbp]
\caption{\label{tbl:AFQMC-chol-cutoff}
GTO-AFQMC total energies
for
several values of the mCD threshold parameter $\delta$
for {\CaHHHH} ($Z = 2.3$\,{\AA} and \mbox{$\dHH = 0.7682$\,{\AA}}; see text),
using the cc-pVTZ basis ($M = 155$).
A fixed Trotter time step $\tau = 0.01 \Eh^{-1}$ was used for the all the
calculations.
The total energy $\EDD$, obtained from direct diagonalization of
$V_{\mu\nu}$, is presented for comparison; the
eigenvalue cutoff is also shown.
$N_\gamma$ is the corresponding number of auxiliary-fields.
A full rank, symmetric, positive definite $V_{\mu\nu}$ matrix
would have required $155^2=24 025$ Cholesky vectors. All energies are in $\Eh$.
}
\begin{ruledtabular}
\begin{tabular}{lrr}
\multicolumn{3}{c}{Direct diagonalization} \\
%&Direct diagonalization&\\
%eigenvalue cutoff ($\Eh$)        & Total energy ($\Eh$)  &  $N_\gamma$ \\  % $\Nfld$  \\
$V_{\mu\nu}$ eigenvalue cutoff & \multicolumn{1}{c}{$\EDD$} &  $N_\gamma$ \\
\hline
\vspace{-0.8em}
\\
$\lambda_\gamma > 2 \times 10^{-8}$
                                 & $-681.42990(20)$      &  $2280$   \\
\hline
\hline
\vspace{-0.5em}
\\
\multicolumn{3}{c}{Modified CD} \\
%Cholesky $\delta$                & Total energy ($\Eh$)  & $N_\mathrm{CD}$  \\ % $\Nfld$  \\
Cholesky $\delta$                & \multicolumn{1}{c}{\EmCD} & $N_\mathrm{CD}$  \\ % $\Nfld$  \\
\hline
\vspace{-0.8em}
\\
$10^{-8}$                        & $-681.43007(15)$      &  $1727$   \\
$10^{-6}$                        & $-681.43003(17)$      &  $1120$   \\
$10^{-5}$                        & $-681.42988(20)$      &  $850$    \\
$10^{-4}$                        & $-681.42977(18)$      &  $643$    \\
$10^{-3}$                        & $-681.42932(19)$      &  $511$    \\
$2 \times 10^{-3}$               & $-681.42991(19)$      &  $468$    \\
$3 \times 10^{-3}$               & $-681.43102(18)$      &  $436$    \\
$3.5 \times 10^{-3}$             & $-681.42958(19)$      &  $425$    \\
$4 \times 10^{-3}$               & $-681.39679(19)$      &  $403$    \\
$5 \times 10^{-3}$               & $-681.39609(17)$      &  $379$    \\
\end{tabular}
\end{ruledtabular}
\end{table}
Figure~\ref{fig:AFQMC-chol-cutoff} plots
the error $\EmCD(\delta)-\EDD$
from Table~\ref{tbl:AFQMC-chol-cutoff}.
For comparison, the corresponding error in the UHF
variational energy, computed using the same set of Cholesky vectors, is also shown
and is seen to correlate very well with the AFQMC energies.

\begin{figure}%[htbp]
\includegraphics[scale=0.31]{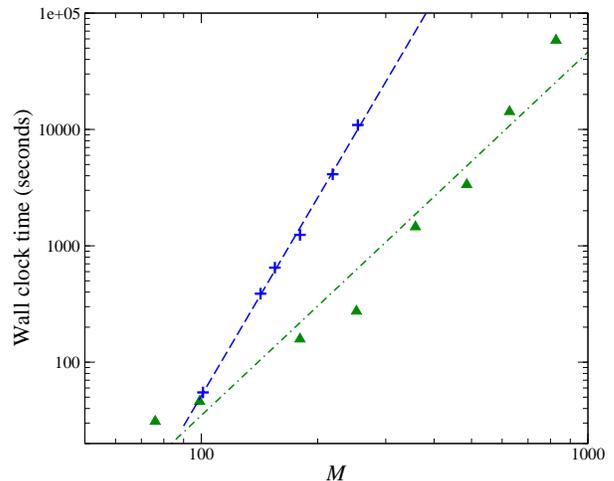}
\caption{\label{fig:diag-vs-chol-timing}
(Color online)
Log-log plots of wall clock times (seconds) vs. basis size $M$, comparing
DD (\textcolor{blue}{$\pmb{+}$}) and
mCD (\textcolor{xmgrace-green4}{$\blacktriangle$})
of the ERI supermatrix $V_{\mu\nu}$.
%The horizontal axis is the number of basis functions, $M$.
%The order of the $V_{\mu\nu}$ supermatrix is $M(M+1)/2$.
The DD is carried out
with 8-thread OpenMP (see text).
The mCD timing is obtained from a locally modified MPQC\cite{MPQC-2.3.1} program,
for fixed $\delta = 10^{-6}$.
No multithreading is used in the mCD calculations.
The dashed (dash-dot) lines are linear regressions.
The DD slope $\sim M^{5.7}$ is consistent with the expected $M^6$ scaling, while
mCD scales as $\sim M^{3.1}$.
}
\end{figure}

% Timing issues
Figure~\ref{fig:diag-vs-chol-timing}
illustrates the relative timing of DD vs.~mCD procedures.
Computations were carried out on 64-bit AMD Opteron
multi-core processors with speeds ranging from 2.2 GHz up to 3 GHz.
\footnote{
    The DD procedure uses the \texttt{dsyev} LAPACK routine
    provided by the Intel Math Kernel Library (MKL).
    Each DD run utilizes eight cores using OpenMP.
    The DD timing only covers the wall clock time required to
    perform the diagonalization of $V_{\mu\nu}$ supermatrix itself.
    The mCD calculations were done using a locally modified MPQC code,
    in which
    the necessary $V_{\mu\nu}$ matrix elements were computed on-the-fly.
    Our reported mCD timings therefore include the time to
    compute these matrix elements.
}
As expected DD times scale as
$M^6$.
The recursive mCD algorithm with prescreening scales only as $\sim M^3$.
%At $M = 180$, for example, the CD procedure beats the DD compute time
%by nearly one order of magnitude.
For the largest ($M = 827$) basis reported in this paper,
mCD required less than six hours on a single core, while DD
would have taken more than 92 days and nearly 1 TB of memory
on the same computer.
%This dramatic example demonstrates the clear advantage of using CD
%in reducing the precomputation time to generate
%the HS form of the two-body operator in \Eq{eq:HS-xform-noFB}.
%

Our current mCD implementation still has much room for improvement for application to larger systems. The
mCD for the preprocessing HS decomposition has not been parallelized. The sparsity of the Cholesky vectors has also not
been exploited for the actual GTO-AFQMC calculations. The decomposition of the  necessary $V_{\mu\nu}$ matrix elements scales
as $\Order{M^3}$ in computer time and the memory required to store the Cholesky vectors also currently scales as $\Order{M^3}$.
Fully exploiting the sparse structure of the Cholesky vectors would
further reduce the memory requirement to $\Order{M^2}$ asymptotically
for very large molecules.\cite{Koch2003}

\subsection{{\CaHHHH} calculation details}
\label{sec:details}

%Allude to discussion for the reason we use the *CV* basis set, etc.

\begin{figure}[htbp]
\includegraphics[scale=0.31]{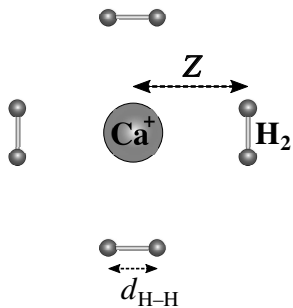}
\caption{\label{fig:Ca4H2}
An illustration of the {\CaHHHH} model chemistry, containing
one {\Caplus} surrounded by four hydrogen molecules in a $D_{4h}$ symmetric
configuration.
%only symmetric dissocation is considered in this work.
%The symmetry group of this ``molecule'' is $D_{4h}$.
}
\end{figure}

As a model of the {\Caplus}
binding site in a hydrogen storage
system,
the PEC for symmetric dissociation of {\CaHHHH}
was calculated as a function of
the {\CaH} lateral distance $Z$ (see Fig.~\ref{fig:Ca4H2}).
For each value of $Z$, the H--H distance $d_\mathrm{H-H}$ is optimized
using MP2 with the cc-pCVTZ basis.
The UHF wave function is used as the trial wave function $\PsiT$.
GTO-AFQMC calculations were carried out using
the correlation-consistent core-valence (cc-pCV$x$Z)
basis set family for Ca
\cite{Peterson2002,Koput2002}
and cc-pV$x$Z for the H atoms.
(This joint basis is subsequently designated as ``cc-pCV$x$Z''.)
For some selected geometries,
calculations were also carried out using a second basis set family denoted ``aug-cc-pCV$x$Z'',
which comprises
the cc-pCV$x$Z functions for Ca
and the diffuse aug-cc-pV$x$Z functions for the H atoms.

Extensive basis-set extrapolation tests were carried out to
eliminate the errors from the use of finite
GTO basis sets. Such tests have
not been systematically reported in previous studies of the system.
The best CBS extrapolation procedure
separately treats
the HF and correlation energies,
\eqsl
{
    \EHF(\infty)
&   \approx
    \EHF(x) - B e^{-\alpha x}
    \,,
    \label{eq:HF-CBS}
\\
    \Ecorr(\infty)
&   \approx
    \Ecorr(x) - \frac{C}{x^3}
    \,,
    \label{eq:corr-CBS}
}
since correlation energy convergence is much slower
than HF.
\cite{Muller2006_basis,Feller1993,Helgaker1997,Halkier1999}
Here $x$ is the correlation consistent basis cardinal number.
The HF CBS extrapolation requires calculations with a minimum of three basis
sets,
while
the correlation energy CBS extrapolation requires at least two.

We examined errors from the Trotter time step $\tau$ on basis set extrapolation in GTO-AFQMC.
With core-valence basis sets, the absolute energy can vary significantly.
At the 5Z level, for example, the error in the absolute energy is
as large as $20$ {\mEh} for $\tau = 0.01 {\Eh}$.
Binding energies, however, are less sensitive,
since the $\tau \to 0$ extrapolation slope for a given basis set was found to be
insensitive to the geometry of the system.
Binding energies were therefore obtained using a finite time step of
$\tau = 0.01 \Eh^{-1}$.
The validity of this approach was verified by
computing energy differences obtained with separate $\tau \to 0$ extrapolations
at
representative geometries.
For the valence-only basis sets, the total energies change insignificantly
under $\tau \to 0$ extrapolation, and the
calculations are always reported at
$\tau = 0.01 \Eh^{-1}$.

The PEC for the {\CaHHHH} binding energy is given by
\eql{eq:Ebind-def}
{
    \Ebind(Z)
    \equiv
    E(Z) - \Efrag
    \,,
}
where $\Efrag$ is the fragment total energy,
\eql{eq:Efrag-def}
{
    \Efrag
    \equiv
    E_{\Caplus} + 4  E_{\Htwo}
    \,.
}
In our CBS extrapolation it is convenient to further decompose the binding
energy into its HF and correlation contributions,
\eql{eq:Ebind-decomp}
{
    \Ebind(Z)
  = \EbindHF(Z) + \Ebindcorr(Z)
    \,,
}
In this work the HF total energies are extrapolated using
$x \in \{ 3, 4, 5 \}$  in $\EbindHF(Z)$.
%For the correlation part,
The ansatz in \Eq{eq:corr-CBS} implies that
$\Ebindcorr(Z)$ varies linearly with respect to $x^{-3}$,
\eql{eq:Ebind-corr-cbs}
{
    \Ebindcorr(Z,\infty)
&   \approx
    \Ebindcorr(Z,x)
  - \frac{C'(Z)}{x^3}
\\
&   \equiv
    \Ebindcorr(Z,x)
  + \Delta\Ebindcorr(Z,x)
    \,.
}
$C'(Z)$ is a $Z$-dependent CBS coefficient.
Ideally we would perform AFQMC calculations at all geometries with
two or more basis sets
to obtain both $\Ebindcorr(Z,\infty)$ and $C'(Z)$ directly.
Such a calculation would be
unnecessarily expensive, especially at the
largest cc-pCV5Z basis level ($M = 627$).
Instead,
AFQMC CBS parameters were obtained at several representative geometries ($Z$)
using both the $x \in \{3,4\}$ and the $x \in \{3,4,5\}$ series.
We then adopt the following strategy to parametrize
$C'_{\textrm{AFQMC}}(Z)$
and estimate the final PEC.
We compute the AFQMC finite-basis $\Ebindcorr(Z,x)$ with the
cc-pCVTZ basis set ($x = 3$).
We also assume that, across all $Z$ values,
the ratio of correlation energies recovered by AFQMC and MP2
is approximately constant and given by the parameter $\rho$.
The AFQMC CBS correction term can therefore be approximated
by scaling the MP2 CBS correction term,
\eql{eq:DEbind-corr-scaled}
{
    \Delta\Ebindcorr(Z,x,\mathrm{AFQMC})
    \approx
    \rho \,
    \Delta\Ebindcorr(Z,x,\mathrm{MP2})
    \,,
}
or, equivalently,
\eql{eq:DEbind-corr-scaled2}
{
    C'_{\mathrm{AFQMC}}(Z)
    \approx
    \rho \,
    C'_{\mathrm{MP2}}(Z)
    \,.
}
The MP2 energies were computed using the cc-pCVTZ and cc-pCVQZ basis sets
to obtain $C'_{\mathrm{MP2}}(Z)$, which determines $\rho$
and the final CBS estimate for the complete AFQMC PEC.
We verified that this MP2-assisted approach accurately reproduced
direct AFQMC
extrapolations at selected geometries to within statistical errors.

Spectroscopic constants associated with
the computed PEC were obtained from Morse fits,
\eql{eq:morse}
{
    \Ebind(Z)
  = E_0
  + \frac{k}{2 a^2}
    \left[
        1 - e^{-a (Z - Z_0)}
    \right]^2
   \, ,
}
where $E_0$ is the well depth minimum, $Z_0$ the location of the well
minimum, and $k$ is the one-dimensional harmonic spring constant.

\section{Results}
\label{sec:results}

\begin{figure}[htbp]
\includegraphics[scale=0.31]{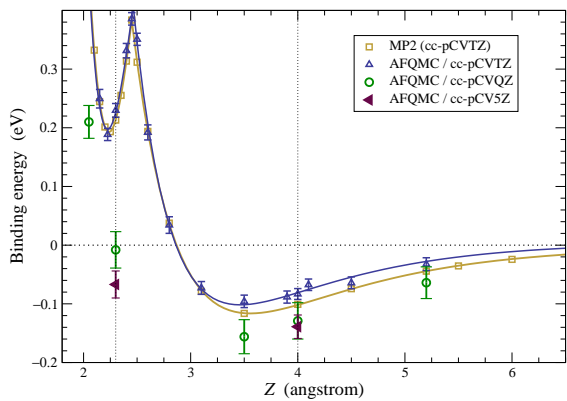}%
\caption{\label{fig:Ca4H2-EOS-cc-pCVTZ}
(Color online)
%Upper panel:
All-electron GTO-AFQMC and MP2 PEC of the {\CaHHHH} symmetric dissociation
as a function of $Z$ ({\Caplus -- \Htwo} separation distance)
at the cc-pCVTZ basis level.
%The hydrogen bond length $\dHH$ is optimized using MP2/cc-pCVTZ
%at every $Z$.
The solid lines are from separate Morse fits for the inner and outer regions.
Also shown
at selected $Z$ are GTO-AFQMC results from larger
cc-pCVQZ and cc-pCV5Z basis sets.
Vertical lines at $Z = 2.3$ {\AA} and $Z = 4.0$ {\AA}
are
a guide to the eye.
}
\end{figure}

In modeling the binding of {\Htwo} molecules onto dispersed calcium centers,
the convergence with respect to basis set is very delicate.
This is illustrated by
the all-electron {\CaHHHH} PECs in
Fig.~\ref{fig:Ca4H2-EOS-cc-pCVTZ}.
The solid lines are Morse fits to GTO-AFQMC and MP2
binding energies calculated with the cc-pCVTZ
basis set (see \Sec{sec:details} for computational details).
The symmetric dissociation PECs are seen to
exhibit a double-well structure.
AFQMC results using larger
(cc-pCVQZ and cc-pCV5Z) basis sets are shown at a few selected $Z$.
The inner and outer wells show dramatically different rates of basis set convergence.
% no paragraph break
% no paragraph break
The inner well
does not exhibit binding at the cc-pCVTZ level,
being $\sim 0.2$\,eV higher than the dissociation limit.
The outer van der Walls well is bound
($\sim -0.1$\,eV at $Z_0 \sim 3.5$\,{\AA}).
As the basis size is increased,
the inner well changes character, becoming $\sim 0$\,eV with  cc-pCVQZ
and then bound by $\sim -0.1$\,eV at the cc-pCV5Z level.
By contrast, the outer well binding energy is already converged
at cc-pCVQZ basis level.
Thus careful extrapolation with the basis sets is required in these systems
to reach the CBS limit.

%\subsection{Convergence to the complete basis limit}
%== \subsection{Extrapolation to the CBS limit}

%\label{sec:basis-extrap}

\begin{figure}[tbp]
\includegraphics[scale=0.31]{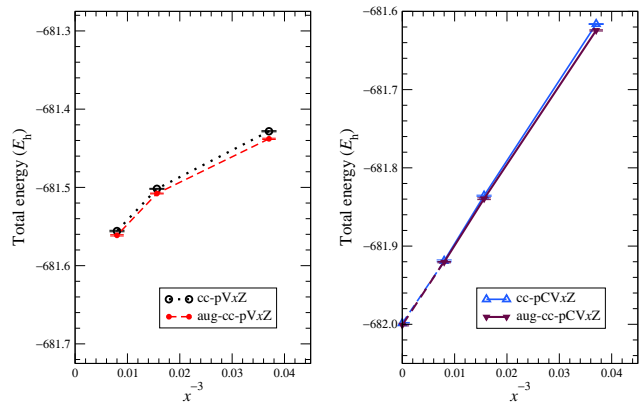}
\caption{\label{fig:Ca4H2-Etot-Z23-extrap-CBS}
(Color online)
Basis set convergence of GTO-AFQMC {\CaHHHH} total energies for
$Z = 2.3$\,{\AA} (near the inner well minimum).
Energies are plotted as a function of $x^{-3}$,
where $x \in \{3,4,5\}$ is the correlation consistent basis cardinal number.
Left panel: valence-only cc-pV$x$Z and aug-cc-pV$x$Z; right panel: core-valence cc-pCV$x$Z and
aug-cc-pCV$x$Z.
}
\end{figure}

We now
show that the basis set must well represent the semicore Ca $3s$ and $3p$ states for
accurate extrapolation.
Figure~\ref{fig:Ca4H2-Etot-Z23-extrap-CBS}
plots the total energies of {\CaHHHH} for
correlated valence basis sets (cc-pV$x$Z and aug-cc-pV$x$Z) and correlated core-valence
basis sets (cc-pCV$x$Z and aug-cc-pCV$x$Z), as a function of basis cardinal number [$x=(3,4,5)$; see Eq.~(\ref{eq:corr-CBS})].
The calculations are for $Z = 2.3$\,{\AA}, which is near the inner well minimum.
With the core-valence basis sets,
the
energies show linear dependence on
$x^{-3}$, consistent with the ansatz in Eq.~(\ref{eq:corr-CBS}),
while the valence-only
series do not.
For reference and benchmark purposes,
selected GTO-AFQMC total energies are tabulated in
Table~\ref{tbl:AFQMC-chol-large-N}.
\begin{table*}[!btp]
\newcommand\ctr[2][1] {\multicolumn{#1}{c}{#2}}
\renewcommand\ctr[2][1] {#2}
\caption{\label{tbl:AFQMC-chol-large-N}
All electron GTO-AFQMC {\CaHHHH} total energies (in {\Eh})
for two geometries and two correlation-consistent core-valence basis sets.
``Inner well'' corresponds to
$Z=2.3$\,{\AA} and $\dHH = 0.7682$\,{\AA};
 ``outer well'' corresponds to
$Z=4.0$\,{\AA} and $\dHH=0.7362$\,{\AA}.
The total energies of the isolated {\Caplus} and {\Htwo} fragments
are also shown.
The mCD method
with $\delta = 10^{-6}\,\Eh$ is used, unless otherwise noted.
In all cases, bias from mCD is much smaller than the AFQMC statistical error,
which are on the last two digits and are indicated in parentheses.
All energies were extrapolated to the $\tau \to 0$ limit.
CBS indicates the complete basis set extrapolated limit.
}
%\begin{ruledtabular}
\begin{tabular}{lcclclclcc}
\hline
\hline
{\qquad\qquad\quad}
             &          && \multicolumn{1}{c}{inner well}
                        && \multicolumn{1}{c}{outer well}
                        && \multicolumn{1}{c}{\Caplus}
                        && \multicolumn{1}{c}{\Htwo} \\
Basis set    &    $M$   && \multicolumn{1}{c}{$Z=2.3$\,\AA}
                        && \multicolumn{1}{c}{$Z=4.0$\,\AA}
                        &&
                        && \multicolumn{1}{c}{$\dHH = 0.7362$\,\AA} \\
\hline
\vspace{-0.5em}
\\
\multicolumn{7}{l}{Basis family: cc-pCV$x$Z} \\
$x = 3$      &   $180$  && $-681.61620(61)$
                        && $-681.62600(71)$
                        && $-676.93411(28)$
                        && $-1.17257(11)$ \\
$x = 4$      &   $358$  && $-681.83561(74)$\footnote{$\delta = 10^{-5}\, \Eh$}
                        && $-681.84005(76)$\textsuperscript{{a}}
                        && $-677.13994(45)$
                        && $-1.17384(18)$ \\
$x = 5$      &   $627$  && $-681.9189(12)$\textsuperscript{{a}}
                        && $-681.92106(77)$\textsuperscript{{a}}
                        && $-677.21997(52)$
                        && $-1.17412(15)$ \\
CBS          & $\infty$ && $-681.9976(10)$
                        && $-681.99954(84)$
                        && $-677.29517(52)$
                        && $-1.17447(17)$ \\
\hline
\vspace{-0.5em}
\\
\multicolumn{7}{l}{Basis family: aug-cc-pCV$x$Z} \\
$x = 3$      &   $252$  && $-681.62437(79)$
                        && $-681.63180(89)$
                        &&
                        && {\hspace{0.55em}}$-1.172826(83)$ \\
$x = 4$      &   $486$  && $-681.8395(11)$
                        && $-681.84151(71)$
                        &&
                        && $-1.17397(22)$\\
$x = 5$      &   $827$  && $-681.92056(80)$\footnote{$\delta = 10^{-4}\, \Eh$}
                        && $-681.9213(10)$\textsuperscript{b}
                        &&
                        && $-1.17434(12)$\\
CBS          & $\infty$ && $-682.00018(94)$
                        && $-681.99718(99)$
                        &&
                        && $-1.17466(14)$ \\
\hline
\hline
\end{tabular}
%\end{ruledtabular}
\end{table*}
% Points in showing this table:
% - show the actual # of fields
% - show how large we can do calculation
% - show how energy converges vs x
% - show
% No paragraph break
\begin{figure}[htbp]
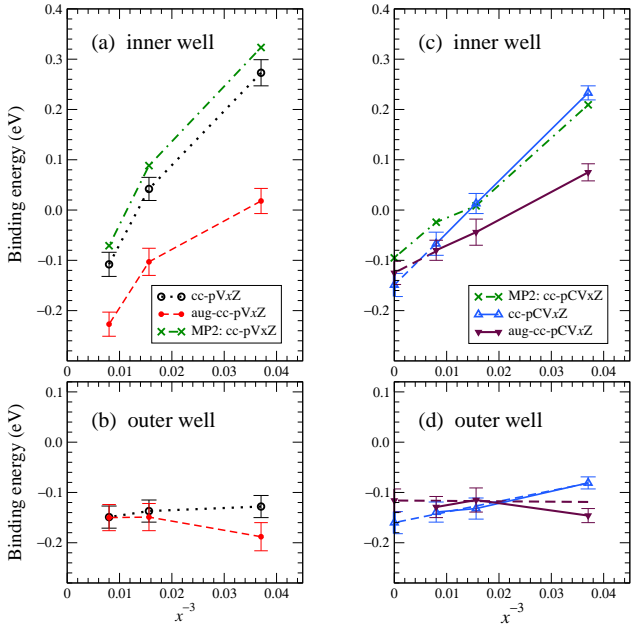

\includegraphics[scale=0.31]{\FigureFile{Ca4H2-Ebind-Z23-extrap-CBS}}
\\
\includegraphics[scale=0.31]{\FigureFile{Ca4H2-Ebind-Z40-extrap-CBS}}
\caption{\label{fig:Ca4H2-Ebind-Z23-Z40-extrap-CBS}
(Color online)
Top panels: The binding energy of {\CaHHHH} system near the inner well minimum
($Z = 2.3$\,{\AA}, $\dHH = 0.7682$\,{\AA})
plotted against $(x^{-3})$,
where $x$ is the correlation consistent basis cardinal number.
Bottom panels: The binding energy near the outer well minimum
($Z = 4.0$\,{\AA}, $\dHH = 0.7362$\,{\AA}).
}
\end{figure}

% start new paragraph:
The importance of proper core-valence treatment is also evident in the binding energies, as shown by
Fig.~\ref{fig:Ca4H2-Ebind-Z23-Z40-extrap-CBS}, %(a)--(d),
where
results for both the inner- and outer-well regions are shown. For comparison, MP2 results are also shown for the inner-well region.
The magnitude of core-valence effects is clearly larger in the inner-well region.
This is due to the Kubas interaction,\cite{Kubas2001,Kubas2007} involving Ca($3d$) states.
Since the spatial extent of the semicore Ca($3s$,$3p$) is similar to the Ca($3d$) states, core-valence effects
are magnified in this region.
The CBS extrapolation lowers
the inner well minimum by nearly $0.4$ eV compared
to the cc-pCVTZ basis results.
By contrast, the outer well depth
is lowered by only  $< 0.1$ eV.
Even at the cc-pCV5Z level the binding energy is still
$\sim 0.05$ eV higher than the
CBS limit for the inner well, while at the the outer well it has long converged.
Figure~\ref{fig:Ca4H2-Ebind-Z23-Z40-extrap-CBS} also shows
that the CBS limit is relatively insensitive
to the use of diffuse "aug" functions for the H atoms.
These results emphasize that it is necessary to use larger core-valence correlation-consistent basis sets in many-body
calculations for such weakly bound hydrogen storage systems with
binding-site atoms containing semicore states.

In Figs.~\ref{fig:Ca4H2-Etot-Z23-extrap-CBS} and
\ref{fig:Ca4H2-Ebind-Z23-Z40-extrap-CBS}
we see that, with the cc-pCV$x$Z basis sets, even the total and binding energies
follow the $x^{-3}$ scaling to a very good degree,
which is the form for correlation energies, as shown in Eq.~(\ref{eq:corr-CBS}).
This is so because the HF energy converges rapidly with the basis size,
as indicated by \Eq{eq:HF-CBS}.
For example, the HF energy
changes by roughly $-4\,\mEh$
from cc-pCVTZ to the CBS limit across different $Z$ values,
in contrast to a change in correlation energy of almost $-400\,\mEh$.
Similarly, the corresponding HF binding energy
changes only by about
$-1\,\mEh$ ($-0.03$ eV).
Thus, using the procedure described in \Sec{sec:details}, the many-body results with
core-valence correlation-consistent basis sets can be extrapolated to the CBS limit
in a very robust fashion.

\begin{table}
\caption{\label{tbl:welldepth}
Binding energies $\Ebind{}$ after extrapolation to the CBS limit. The two geometries are the same as
in Table~\ref{tbl:AFQMC-chol-large-N} (inner well at $Z=2.3$\,{\AA} and outer well at
$Z=4.0$\,{\AA}).
The contributions to the AFQMC
binding energy [$\Ebind(Z,\infty)$ on the third row] from HF
and correlation are shown separately in the first two rows. Energies are in eV. AFQMC
statistical errors are shown in parentheses.
}
\begin{ruledtabular}
\begin{tabular}{lll}
                    & Inner well       & Outer well       \\
\hline
\vspace{-0.8em}
\\
%\multicolumn{3}{l}{\textbf{Extrapolation to CBS limit}}   \\
$\EbindHF(Z,\infty)$
                    & $+0.815$    &  $-0.082$             \\
$\Ebindcorr(Z,\infty)$
                    & $-0.954(23)$     &  $-0.072(22)$    \\
$\Ebind(Z,\infty)$
                    & $-0.139(23)$     &  $-0.154(22)$    \\
\end{tabular}
\end{ruledtabular}
\end{table}

Table~\ref{tbl:welldepth} shows the extrapolated binding energy results
at two representative geometries.
HF is qualitatively correct for the outer well, capturing more than half of
the well depth. For the inner well, HF is unbound by a very large amount on the
scale of interest, by more than $0.8$\,eV. Thus, consistent with the discussion above
on the slow convergence of the Kubas complex, the binding of the inner well is dominated by electron correlation
effects, which contribute almost $1$\,eV to the well depth.

%\subsection{Potential energy curve for {\CaHHHH} symmetric dissocation}
%\label{sec:PEC}

\begin{figure}
\includegraphics[scale=0.34]{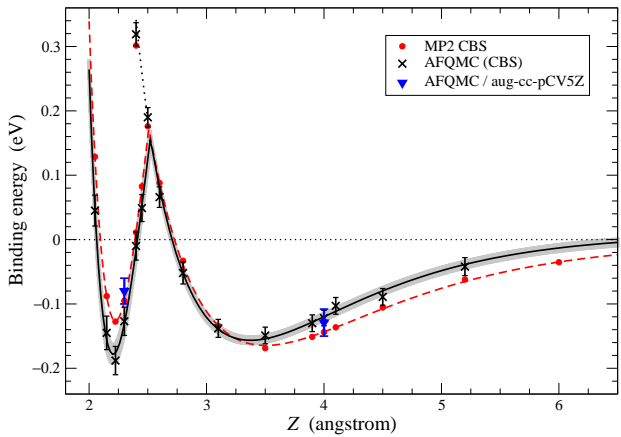}%
\caption{\label{fig:Ca4H2-EOS-CBS}
(Color online)
The PEC of the {\CaHHHH} symmetric dissociation
extrapolated to the CBS limit.
The PEC is shown as a function of {\Caplus -- \Htwo}
separation distance, $Z$.
Morse fits to the AFQMC points are shown as solid curves.
The gray shading provides an estimate of the PEC uncertainties.
Dotted lines shows the continuation of the outer well PEC into
the small $Z$ region.
Morse fits to the MP2 points are shown as dashed curves.
The extrapolation scheme is discussed \Sec{sec:details}.
For comparison, binding energies computed using very large
aug-cc-pCV5Z basis sets are also shown at $Z=2.3$ and $4.0$\,{\AA}.
}
\end{figure}

Figure~\ref{fig:Ca4H2-EOS-CBS} presents the final
 {\CaHHHH} symmetric PEC from GTO-AFQMC calculations,
after extrapolation to the CBS limit.
The GTO-AFQMC binding energies from the  aug-cc-pCV5Z  basis set are shown
at
$Z=2.3$
and $4.0$\,{\AA} to indicate the effect of the extrapolation (recall Fig.~\ref{fig:Ca4H2-EOS-cc-pCVTZ}).
The extrapolation of the AFQMC results was done with the procedure described in
Sec.~\ref{sec:details}, assisted by MP2 CBS corrections.
As mentioned, this procedure was verified by
direct extrapolations of the AFQMC results at two representative geometries
using the $x\in\{3,4,5\}$ series, as shown in Table~\ref{tbl:welldepth}, which gave
completely consistent results.
In addition, at multiple other geometries, extrapolations of the AFQMC results
using the $x\in\{3,4\}$ series were done as further confirmation, which
lead to statistically indistinguishable results at the CBS limit.

The corresponding MP2 results at the CBS limit are also shown.
MP2 is seen to represent the double-well structure reasonably well (Figs.~\ref{fig:Ca4H2-EOS-cc-pCVTZ} and
\ref{fig:Ca4H2-EOS-CBS}). However,
there are some significant deviations from AFQMC.
The MP2 inner well is more shallow:
at the largest basis set (cc-pCV5Z),
the MP2 inner-well binding energy is $\sim 0.05$ eV
smaller than AFQMC.
The MP2 outer well, by contrast, is  deeper and broader than
that of AFQMC.
This is perhaps not surprising, given the much stronger effect of electron correlation
in the inner well region.
We comment more on the MP2 results in Sec.~\ref{sec:discuss}.

The solid lines with gray shading in Fig.~\ref{fig:Ca4H2-EOS-CBS} represent
Morse fits of the GTO-AFQMC binding energies at the CBS limit,
where the width of the shading
indicates the uncertainties in the fits.
As mentioned, the two regions have separate Morse fits.
The fitting uncertainty
in the inner well is $\lesssim 0.02$ eV, and
$\lesssim 0.01$ eV for the outer well.
Spectroscopic constants corresponding to the Morse fitted GTO-AFQMC CBS results are shown in
Table~\ref{tbl:spectros-const}.
The two well minima have comparable well depths within the AFQMC statistical
accuracy.

\begin{table}
\caption{\label{tbl:spectros-const}
Spectroscopic constants associated with the
AFQMC inner and outer wells depicted in \Fig{fig:Ca4H2-EOS-CBS}.
These parameters were obtained by fitting a Morse curve to
the two wells separately.
The error bars shown in parentheses below include both
the fitting and AFQMC statistical uncertainties.
}
\begin{ruledtabular}
\begin{tabular}{lcrr}
Quantity                   & Units        & Inner well    & Outer well    \\
\hline
Well depth minimum $E_0$   & eV           & $-0.178(16)$  & $-0.1566(71)$ \\
Equilibrium distance $Z_0$ & {\AA}        &  $2.205(17)$  &   $3.375(41)$ \\
Spring constant $k$        & eV/{\AA}$^2$ &   $13.0(22)$  &   $0.342(37)$ \\
\end{tabular}
\end{ruledtabular}
\end{table}

\section{Discussion}
\label{sec:discuss}

In the previous section, the  GTO-AFQMC {\CaHHHH} PEC 
was shown to have a double-well structure with
weak binding of four {\Htwo} molecules.
After extrapolation to the CBS limit, 
binding energies of
$-0.178(16)$ and $-0.1566(71)$ eV were found at the respective minima of the inner and outer wells,
essentially the same within statistical errors.
The crossover between the two wells occurs at $Z \sim 2.5$\,{\AA}.
Below we first comment further on possible sources of uncertainty.

A possible source of error in the PEC is the basis set superposition
error (BSSE).
For a finite basis, when two or more fragments are brought together,
the resulting ``molecule'' enjoys additional degrees of freedom not
present in the isolated fragments.
This BSSE results in the artificial enhancement of the predicted binding
energy.
Counterpoise (CP) corrections \cite{Duijneveldt1994}
represent an attempt to
reduce the BSSE.
In the {\CaHHHH} system, the effect of the CP correction on HF energies is
negligible even at the relatively small cc-pCVTZ basis level.
For many-body calculations, we investigated the effect of the CP correction within MP2.
The binding energy at the CBS limit was changed
by $\sim 0.02$ eV for the inner well,
while the correction in the outer well is negligible.
This estimate indicates that the effect of CP correction is
within the statistical error of GTO-AFQMC
and does not change our conclusions.

A second and related possible source of error is the basis set convergence, as
discussed extensively in the previous two sections. This is a system
which is particularly demanding in terms of reaching the CBS limit, as we have already shown.
The consistency
of our various cross-checks suggest that the errors from basis set extrapolation
are captured in the indicated uncertainties in Fig.~\ref{fig:Ca4H2-EOS-CBS} and
in Table~\ref{tbl:spectros-const}.
After we completed this work we were made aware of more recent
core-valence basis sets by Iron and co-workers \cite{Iron2003}.
They have been employed to determine spectroscopic constants for
quantum defect calculations in calcium.\cite{Kay2008,Kay2011}
We have tested these new basis sets,
with and without the additional d functions,
using MP2 on the {\CaHHHH} binding energy at $Z = 2.3$\,{\AA}.
We found that the binding energy results were consistent with those from the
cc-pCV$x$Z basis sets which we have employed in the present paper.
For example, the binding energy at the 5Z level agreed to within $0.003$ eV.

A third possible source of error is that our AFQMC method is not exact.
The phaseless approximation is made to
control the sign/phase problem, and as a result there is a systematic bias. The
ground state energies calculated from the method is not guaranteed to be a variational upper bound.
As mentioned, for a variety of benchmarks and applications, the systematic bias is shown
to be very small, consistently reaching the level of accuracy of CCSD(T). For bond-stretching
and bond-breaking, the method is shown to be more accurate than CCSD(T).
\cite{AlSaidi2006b,AlSaidi2007b,Purwanto2008,Purwanto2009_C2}
For the present
systems, internal checks by varying the form of the trial wave function indicate that
the results are very robust.

%\subsection{Comparison with other correlated methods}
%\label{sec:MP2}

We next compare our results with previous calculations.
As discussed in the previous section, MP2 is seen to describe the system quite well.
Due to its nonperturbative nature,
AFQMC recovers more correlation energy compared to MP2.
For example, at the CBS limit,
the AFQMC {\CaHHHH} total energies are $\sim 60\,\mEh$ lower at all
geometries.
Cancellation of errors greatly reduces the discrepancies in the corresponding binding energies.
The high symmetry and absence of near-degeneracy in the {\CaHHHH} system
also makes it easier for MP2 to perform well.
The fourfold symmetric  {\CaHHHH} system
is in a half-filled, closed-shell
configuration.
Open-shell configurations and near degeneracies in other cases would be more challenging.

The CCSD(T)/CBS results of Ohk and co-workers \cite{Ohk2010}
also predict binding for the both the inner and outer wells.
However, Ohk {\it et al.} used valence-only correlation consistent basis sets, and only up to cc-pVQZ to find the CBS limit.
As shown in the previous section, extrapolations with valence-only basis sets is problematic.
Moreover, their crossover position
is at $Z \sim 2.3$ {\AA}, while our AFQMC and MP2 crossovers are at $Z \sim 2.5$ {\AA}.

In comparing with the DMC results of Bajdich
\textit{et al.}, \cite{Bajdich2010} we note that in general DMC is a highly
accurate many-body method.
Their calculation predicts an inner well that is not bound
($E_0 \sim 0.06$ eV)
and an outer well that is weakly bound
($E_0 \sim -0.07$ eV).
Their crossover position
is more consistent with our PEC than with the results of Ohk {\it et al.} \cite{Ohk2010}
They used $Z = 4.6$ {\AA} as representative of the unbound fragments.
Our result suggests that at $Z = 4.6$ {\AA} the {\CaHHHH} is still weakly
bound with $\Ebind \sim -0.08$ eV.
Correcting the DMC PEC
by this amount
increases the DMC outer well binding strength to agree with AFQMC.
The DMC inner well remains essentially unbound.
The DMC calculations  fixed ${\dHH} = 0.77$ {\AA},
but our test calculations with AFQMC varying $\dHH$ show that
this only has a small effect on the binding energy,
$< 0.01$ eV. Thus, the
most likely cause for the discrepancy on the inner well
would appear to be the fixed-node error in DMC.
The use of only triple-zeta quality basis sets in the trial wave functions of the
DMC calculations may have contributed to increase this error.

\section{Summary}
\label{sec:summary}

%In this paper we present our AFQMC calculations of the PEC
%of the {\CaHHHH} complex as a model of
%{\Htwo} binding site in a hydrogen storage system.
The phaseless auxiliary-field
quantum Monte Carlo method
%[S.~Zhang and H.~Krakauer, Phys.~Rev.~Lett.~\textbf{90}, 136401 (2003);
%W.~A.~Al-Saidi, S.~Zhang and H.~Krakauer, J.~Chem.~Phys.~\textbf{124},
%224101 (2006).]
was used to accurately predict the binding
energy of {\CaHHHH},
a model chemistry that has recently been used
to test the reliability of electronic structure methods for {\Htwo}  binding on dispersed alkaline
earth centers.
A modified Cholesky decomposition
is implemented to realize the
Hubbard-Stratonovich transformation efficiently in AFQMC
with large basis sets, which removes a memory and
computational bottleneck.
We employ the largest correlation consistent Gaussian-type basis sets available,
up to \mbox{cc-pCV5Z} for Ca, 
to accurately extrapolate to the complete basis limit.
The
resulting potential energy curve exhibits binding with a
double-well structure, with
nearly equal binding energy minima of \mbox{$\sim -0.18$ eV}.  %(or $\sim -0.04$ eV/{\Htwo}).
%
%We perform careful basis set extrapolation using the
%correlation-consistent core-valence basis set (cc-pCV$x$Z) series.
%The proper wave function description of physisorption state
We showed that an accurate description of the inner well
Kubas complex requires
the use of the correlation-consistent core-valence basis set series \mbox{cc-pCV$x$Z}.
While the model's binding energy of \mbox{$\sim -0.04$ eV/{\Htwo}}
falls short, in itself, of the targeted optimum design value by more than
a factor of three,
the results are encouraging for further study
of larger, more realistic
models as potential candidates for hydrogen storage media.

The results in this paper demonstrate that
the phaseless AFQMC method  can accurately treat the weakly bound
systems of interest for hydrogen storage.
This is consistent with earlier results on a variety of materials systems.
As GTO-AFQMC continues to improve,
it can be expected to treat larger, more realistic, nanoscale size systems, with the largest available GTO basis sets.
We hope
that the latest results will encourage the integration of the GTO-AFQMC method with quantum chemistry approaches,
as a component of the
community's efforts for improving energy technologies.

\begin{acknowledgments}

The work was supported in part by grants from DOE (DE-SC0001303 and DE-FG05-08OR23340), ONR (N00014-08-1-1235), and NSF (DMR-1006217).
%
% Standard acknowledgment from NCCS.gov website
%
This research used resources of
the Oak Ridge Leadership Computing Facility,
located in the National Center for Computational Sciences
at Oak Ridge National Laboratory,
which is supported by the Office of Science of the DOE
under Contract DE-AC05-00OR22725.
We also acknowledge the
computing support from
the Center for Piezoelectrics by Design.
We are grateful to Garnet Chan, Fred Manby, and Todd Martinez for helpful conversations, and to Eric Walter for many useful discussions
throughout the course of this work.
{W.P.} would like to thank Jeff Hammond for useful discussions
and suggestions.

\end{acknowledgments}

\bibliography{AFQMC-bib-entries,Energy-applications,Miscellaneous}

\end{document}